\documentclass[pra,twocolumn,showpacs,preprintnumbers,amsmath,amssymb]{revtex4}

\usepackage{graphicx}
\usepackage{dcolumn}
\usepackage{bm}

\begin{document}


\title{Thermodynamic Limit for the Ising Model on the Cayley Tree}

\author{Borko D. Sto\v si\' c}
\email{borko@ufpe.br}
\affiliation{
Departamento de F\' \i sica e Matem\' atica, 
Universidade Federal Rural de Pernambuco,\\
Rua Dom Manoel de Medeiros s/n, Dois Irm\~ aos,
52171-900 Recife-PE, Brasil
}
\author{Tatijana Sto\v si\' c}
\affiliation{
Laboratory for Theoretical Physics, 
Institute for Nuclear Sciences, \\
Vin\v ca, P.O. Box 522, YU-11001 Belgrade, Yugoslavia
}
\author{Ivon P. Fittipaldi}
\affiliation{
Minist\' erio da Ci\^ encia e Tecnologia, 
Esplanada dos Minist\'erios, Bloco E, 2$^o$ andar, Sala 215, 70067-900 Bras\' \i lia-DF, Brazil
}

\date{\today}

\begin{abstract}
While the Ising model on the Cayley tree has no spontaneous magnetization at
nonzero temperatures in the thermodynamic limit, we show that finite systems 
of astronomical sizes remain magnetically ordered in a wide temperature range,
if the symmetry is broken by fixing an arbitrary single (bulk or surface) spin.
We compare the behavior of the finite size magnetization of this model
with that of the Ising model on both the Sierpinski Gasket, 
and the one-dimensional linear chain. This comparison reveals the analogy of
the behavior of the present model with the Sierpinski Gasket case.
\end{abstract}

\pacs{05.50.+q, 64.60.Cn, 75.10.Hk}

\maketitle

The zero-field partition function of the Ising model on the Cayley tree is identical
to that of the one-dimensional linear Ising chain \cite{heimburg}, and thus it does not
lead to any singularities of the zero-field thermodynamic response functions.
This model manifests thermodynamic singular behavior only in response to the
magnetic field, and it has been well established \cite{heimburg,matsuda,muller,morita}
that the susceptibility diverges in a wide range of temperatures, while spontaneous 
magnetization remains zero at any finite temperature in the thermodynamic limit.

While the above facts were established three decades ago, the exact analytical
expression for the zero field magnetization has been derived only recently
\cite{melin,stosic}. Similar to the expression for the zero-field
partition function, the obtained
expression for magnetization is equivalent to that of the linear chain, 
except for the fact that the number of particles
is replaced by the tree generation level.
It is shown in this work that this fact, together with fixing an arbitrary single spin, 
leads to magnetic ordering in a
wide temperature, range for systems of astronomical sizes.
This is reminiscent of the effect of
extremely slow decay of correlations found for the Ising model on the
Sierpinski gasket \cite{liu}, where it has been demonstrated that thermodynamic 
limit represents
a poor approximation for systems of both laboratory and astronomical dimensions.
We compare the finite size behavior of zero field magnetization of the Ising
model on the Cayley tree with that of the Sierpinski gasket and the linear chain.

For simplicity we consider here explicitly only 
the Cayley tree with the branching number (coordination number minus one) 
$B=2$, the generalization to arbitrary $B$ being straightforward.
Following Eggarter \cite{egarter}, we further consider a single
$n$-generation branch of a Cayley tree, composed of two
$(n-1)$-generation branches connected to a single initial site, with
the Hamiltonian
\begin{equation}
{\cal H}=-J\sum_{\langle nn\rangle}S_iS_j-H\sum_{i}S_i,\label{one}
\end{equation}
where $J$ is the coupling constant, $H$ is the external magnetic field,
$S_i=\pm 1$ is the spin at site $i$, and $\langle nn\rangle$ denotes
summation over the nearest-neighbor pairs. 
The $n$-generation branch consists of $N_n={2^{n+1}-1}$ spins, the
$0$-generation branch being a single spin.  The recursion relations for
the partition function of any two consecutive generation
branches were found by Eggarter \cite{egarter} to be
\begin{equation}
Z_{n+1}^{\pm}=y^{\pm 1}\left[ x^{\pm 1}Z_{n}^{+} 
+ x^{\mp 1}Z_{n}^{-}\right] ^2,\label{two}
\end{equation}
where $x\equiv \exp(\beta J)$, $y\equiv \exp(\beta H)$, 
$\beta\equiv 1/k_BT$ is the reciprocal of the product of
the Boltzmann constant $k_B$ and temperature $T$,
and
$Z_{n}^{+}$ and $Z_{n}^{-}$ denote the branch partition functions
restricted by fixing the initial spin (connecting the two
($n-1$)-generation branches) 
into the $\{ +\}$ and $\{ -\}$ position, respectively. 

It turns out that the nonlinear coupled recursion relations (\ref{two}) cannot be iterated 
to yield a closed form expression in the general nonzero field case
\cite{heimburg,matsuda,muller,morita,egarter}.
However, only recently the present authors have arrived \cite{stosic} at the exact 
analytical expression for the zero-field magnetization and susceptibility, 
by considering the recursion relations 
for the {\it field derivatives} of the partition function, which can be iterated in 
the limit $H\longrightarrow 0$ to yield corresponding closed form expressions for
systems of arbitrary sizes (therefore also for the thermodynamic limit).
The exact expression for restricted magnetization, for arbitrary tree generation level $n$, 
was found to be \cite{stosic}, 
\begin{equation}
<m>_n^{\pm}=
\pm {\frac {\left (2\,t\right )^{n+1}-1}
{\left ({2}^{n+1}-1\right )\left (2\,t-1\right )}},\label{three}
\end{equation}
where $t\equiv\tanh (\beta J)$.
The same expression was (also only recently) independently obtained 
by M\' elin {\it et al.} \cite{melin}, using a different approach.

It turns out that the zero-field magnetization 
(as well as all the other odd field derivatives of the partition function) 
is identically zero in strictly zero field, as the expressions 
corresponding to the two possible orientations of the initial spin differ only in sign.
However, this is also true for an arbitrary Ising system in (strictly) zero field, 
as every spin configuration has a mirror image (obtained by flipping all the spins) 
with exactly the same energy and inverted sign of the configurational magnetization,
and to remove this degeneracy one first has to break the symmetry.
The usual procedure of breaking the symmetry by retaining an infinitesimal field, then taking
the thermodynamic limit, and finally taking the zero field limit, is not suitable here
because of the fact that susceptibility diverges in a wide temperature region, rather
than in a single (critical) point. 

Instead, here it seems most suitable to break the
symmetry simply by fixing the initial spin in the upward position (therefore disregarding
all configurations corresponding to downward initial spin orientation), 
and in what follows we shall
first examine the analytical results corresponding to this situation.
It can be argued, however, that the initial spin has a privileged role in this lattice,
and we shall here also examine numerically the consequence of fixing an arbitrary spin
(on the surface, or in the bulk), on the thermodynamic behavior of extremely large
(but finite) systems. It will be shown in the rest of this paper that fixing any one
spin, which is equivalent to imposing an infinitely strong local field of infinitesimal
range (acting on a single spin), leads to symmetry breaking, causing
magnetic ordering of systems of astronomical sizes in a wide temperature range.

By fixing the initial spin, one can consider as the order parameter
the restricted magnetization
\begin{equation}
<m>_n^{+}\equiv{\frac{1}{N_n}}\,{1\over Z_n^{+}}
{{\partial Z_{n}^{+}}\over{\partial \beta H}} =
{\frac {\left (2\,t\right )^{n+1}-1}
{\left ({2}^{n+1}-1\right )\left (2\,t-1\right )}}. \label{four}
\end{equation}
As $t<1$ for $T\neq 0$, this expression clearly demonstrates that
there is no magnetic ordering in the thermodynamic limit 
at any nonzero temperature, 
confirming earlier results \cite{heimburg,matsuda,muller,morita}.
However, comparing with the corresponding expression for the one-dimensional Ising
model (which may also be considered as a Cayley tree with the branching number B=1)
\begin{equation}
<m>_n^{+}=
{\frac {\left (t\right )^{n+1}-1}
{\left (n+1\right )\left (t-1\right )}},\label{five}
\end{equation}
one sees that the generation level is found in place of the number of particles
(these two being the same in the case of the chain). If one takes the Avogadro's number
($\sim 10^{23}$), and the estimated number of hadrons in the observable Universe 
($\sim 10^{80}$), as the two basic reference points in our description of the physical
world (as opposed to the mathematical definition of infinity), 
it turns out that the slight difference between
expressions (\ref{four}) and (\ref{five}) leads to strikingly different behavior of the
two systems. More precisely, it will be shown in this work that the 
Cayley tree remains ordered in a wide temperature region for systems by far exceeding,
in number of particles, the size of the observable Universe. This behavior is
reminiscent of that of the Ising model on the Sierpinski Gasket, and may
provide some further insight into the thermodynamics of model systems on 
hierarchical and fractal lattices in general, the latter nowadays being commonly
accepted as frequent representatives of our physical world.

Let us begin this (extreme) finite size analysis by explicitly plotting magnetization 
of the Cayley tree and the Ising linear chain, as given by expressions
(\ref{four}) and (\ref{five}), for various system sizes $N<2^{2^{10}+1}-1\sim 10^{309}$ 
($N=2^{n+1}-1$ for the Cayley tree, and $N=n+1$ for the Ising linear chain), as shown in 
Figs.~1a) and 1b), respectively. 
It may be interesting to mention here that the 
extremely simple form of expression
(\ref{five}) is deceptive in the sense that it requires special numerical 
handling when such large system sizes are considered 
(for $N=2^{2^{10}+1}-1$ two thousand digit precision was
needed by the Maple VI software for algebraic
manipulation, for numerical calculation of magnetization at low temperatures).

\begin{figure}[htp]
\includegraphics[width=3.5in]{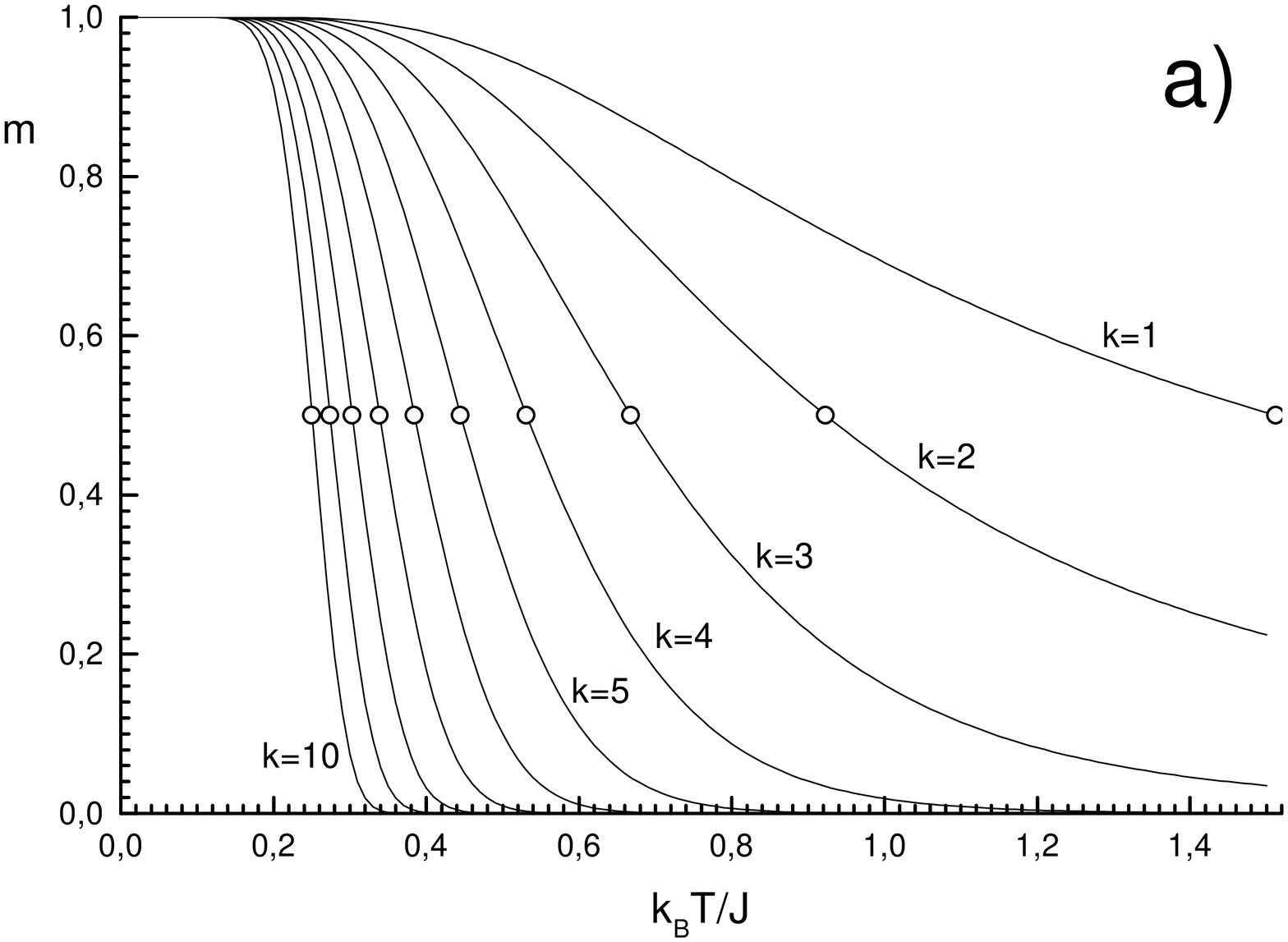}
\includegraphics[width=3.5in]{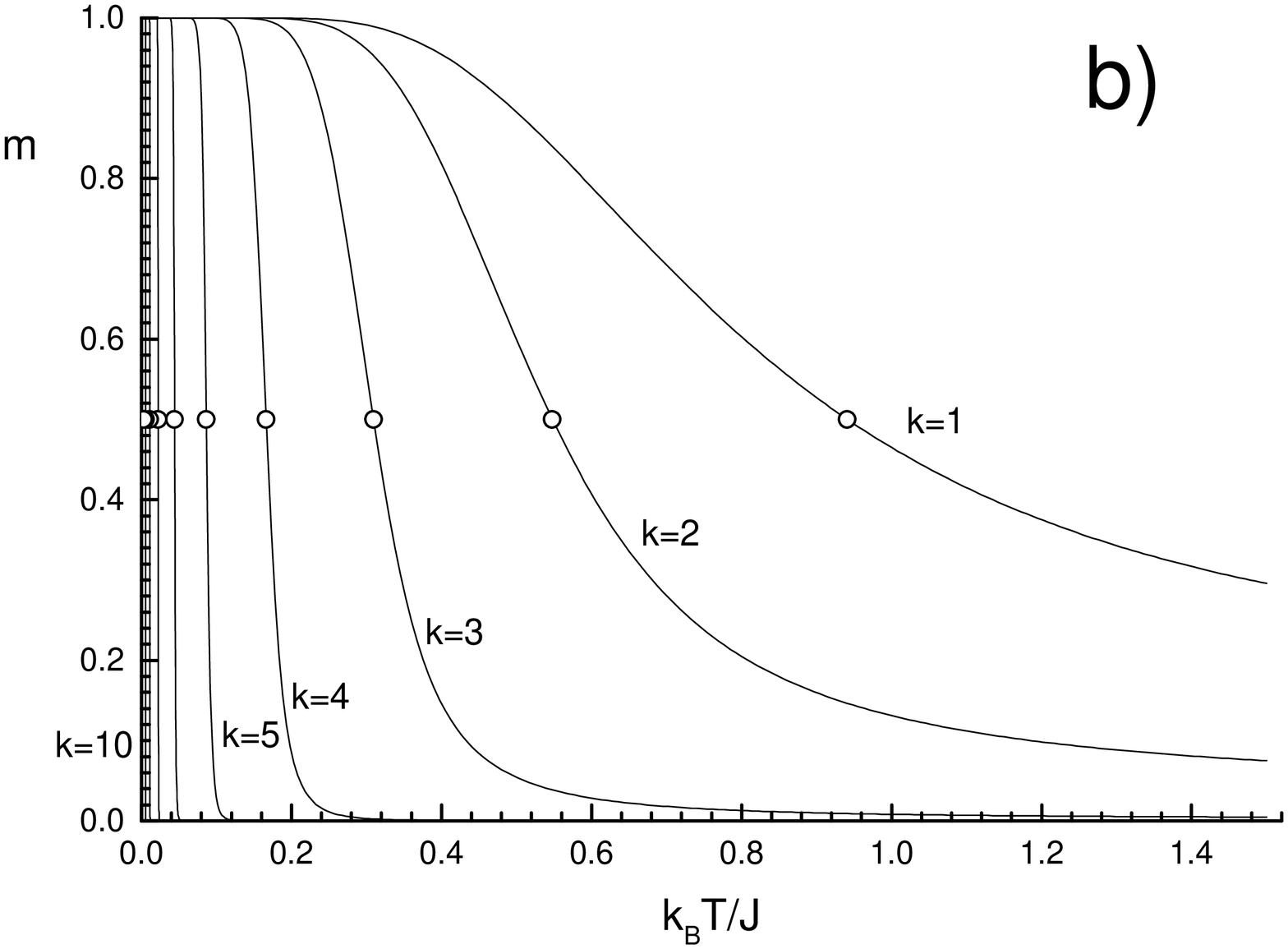}
\caption{ 
Magnetization of the a) Ising model on the Cayley tree, and
the b) the Ising linear chain, for systems with $N=2^{2^k+1}-1$ particles,
with $k=1,2,...,10$. Open circles represent the points where
magnetization falls to half of it's maximum value.
}
\label{fig1}
\end{figure}

One can see from Fig.~1a that even for these inconceivably massive systems (with more then
200 orders of magnitude more particles than the estimated number of hadrons in
the observable Universe), magnetization of the Cayley tree remains substantially removed
from its limiting value of zero, while from Fig.~1b it is seen that 
the magnetization of the Ising linear chain for the
same system sizes far more rapidly collapses to zero, 
and cannot be distinguished from its limiting zero value on the scale of the graph.

We have also performed similar calculations for the 
Ising model on the Sierpinski Gasket, for which it was established
by Liu \cite{liu} that the decay of correlations is extremely slow, 
and the mean square magnetization $\left<m^2\right>_n$ approaches its limiting 
zero value in such a slow fashion that the thermodynamic limit represents a 
poor approximation for both laboratory size systems, 
and systems of
astronomic dimensions. 
In this case the lattice at stage $n$ is constructed by assembling three triangular 
structures of stage $n-1$, and rather then considering $\left<m^2\right>_n$ as
was done in reference \cite{liu}, one
can break the symmetry by fixing the three vortex spins in the current construction 
stage, and consider the corresponding restricted magnetization $<m>_n^{+++}$.
In Fig.~2 we show the comparison of the
points $T_m$, where the restricted magnetization falls to half of its maximum value,
for the Ising model on the Sierpinski gasket, the Cayley tree, and the linear chain.

\begin{figure}[htp]
\includegraphics[width=3.7in]{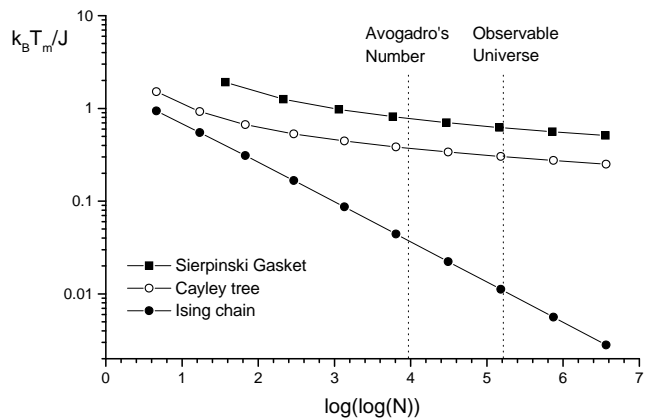}
\caption{
Points where magnetization falls to half of its maximum value,
for the Ising system on the Sierpinski gasket (full squares), 
Cayley tree (open circles) and
the linear chain (full circles), as a function of the number of particles. 
The two vertical dotted lines
indicate system sizes corresponding to the Avogadro's number, and the
estimated number of hadrons in the observable Universe, respectively.
}
\label{fig2}
\end{figure}

It is seen that the magnetization of the Cayley tree 
demonstrates behavior similar to the Sierpinski gasket,
with an extremely slow decay of ordering with the number of particles 
(observe the double logarithmic scale on the abscissa),
in complete contrast with the linear Ising chain.
It should be noted that while both in the case of the
linear chain and the Sierpinski gasket the considered system sizes also imply astronomic
linear dimensions, in the case of the Cayley tree the system linear dimension remains 
miniscule (for the largest considered systems of $N=2^{2^{10}+1}-1$ particles, the largest 
distance between any two spins remains only $2^{10}=1024$ lattice units). In the
case of the Cayley tree one cannot therefore talk of ``long range correlations",
rather, the extremely slow decay of magnetic ordering with temperature may be 
attributed to the short system span and the fact that every spin ``sees" a huge
number of other  spins a short distance away (perhaps a term ``wide range 
correlations" would be more appropriate). In the case of other fractal and
hierarchical lattices, it remains to be seen to what extent the relationship 
between ``long range" and ``wide range" correlations may contribute to the magnetic 
ordering of finite systems. In particular, the question remains whether real porous
samples with magnetic properties that may well be microscopically explained by model 
systems which do not display spontaneous magnetization in the thermodynamic limit, 
still may exhibit magnetic ordering at measurable temperatures, due to both 
their intricate structure and their finite size.

As mentioned earlier, it can be argued that the initial spin occupies a privileged
role in the Cayley tree, and here we address the question of whether fixing any other
spin is sufficient to break the symmetry, and cause ordering of massive systems in a
wide temperature range. Let us therefore consider the situation of building the Cayley
tree in the regular fashion (using recursion relations (\ref{two})) up to level $p$,
fixing a single spin at that level, and then applying modified recursion 
relations (corresponding to asymmetric restricted partition functions) up to level $q$.
From here on, we shall term the branch containing the fixed spin as the ``left" branch,
denoting the corresponding restricted partition functions by $w_n^{\pm}$, in parallel with
the regular ``right" branches (that do not contain any fixed spins), with the ``regular"
restricted partition functions $Z_n^{\pm}$.
Choosing $p=0$ corresponds to a single surface spin being fixed, $p=q$ corresponds to
fixing the initial spin, and $0<p<q$ corresponds to fixing a single bulk spin at level $p$.

Formally differentiating equation (\ref{two}) with respect to field, the recursion 
relations for the field derivatives of the restricted partition functions up to level $p$ are
found to be \cite{stosic}
\begin{eqnarray}
&&{\frac{\partial Z_{n+1}^{\pm}}{\partial h}}=
y^{\pm 1} \left [ \vbox to 14pt{}
\pm \left (Z^{+}_{n}{x^{\pm 1}}+Z^{-}_{n}{x^{\mp 1}}\right )^{2}+
\right.\nonumber\\
&&\left.+2 \left (Z^{+}_{n}{x^{\pm 1}}+Z^{-}_{n}{x^{\mp 1}}\right )
\left (
{\frac{\partial Z_{n}^{+}}{\partial h}}{x^{\pm 1}}+
{\frac{\partial Z_{n}^{-}}{\partial h}}{x^{\mp 1}}
\right ) \right ], 
\label{six}
\end{eqnarray}
where we have used notation $h\equiv \beta H$ for the reduced field.

At all following levels (including $p+1$) the ``right" partition functions
will differ from the ``left" partition functions, and the general recursion
relation is given by
\begin{equation}
w_{n+1}^{\pm}=
y^{\pm 1}\left[ \left( w_{n}^{+}x^{\pm 1}+ w_{n}^{-}x^{\mp 1}\right)
\left(Z_{n}^{+}x^{\pm 1}+Z_{n}^{-}x^{\mp 1}\right)
 \right],\label{seven}
\end{equation}
where for the derivatives we have
\begin{widetext}
\begin{eqnarray}
{\frac{\partial w_{n+1}^{\pm}}{\partial h}}=&&
y^{\pm 1} \left [ 
\pm \left( w_{n}^{+}x^{\pm 1}+ w_{n}^{-}x^{\mp 1}\right)
\left(Z_{n}^{+}x^{\pm 1}+Z_{n}^{-}x^{\mp 1}\right)
+\left({\frac{\partial w_{n}^{+}}{\partial h}}x^{\pm 1}+
{\frac{\partial w_{n}^{-}}{\partial h}}x^{\mp 1}\right)
\left( Z_{n}^{+}x^{\pm 1}+ Z_{n}^{-}x^{\mp 1}\right)
\right.\nonumber\\
&&\left.
+\left( w_{n}^{+}x^{\pm 1}+ w_{n}^{-}x^{\mp 1}\right)
\left({\frac{\partial Z_{n}^{+}}{\partial h}}x^{\pm 1}+
{\frac{\partial Z_{n}^{-}}{\partial h}}x^{\mp 1}\right)
\right ].\nonumber\\ 
\label{eight}
\end{eqnarray}
\end{widetext}

We now take one ``left" branch of level $p$ 
(by setting $w_{p}^{+}=Z_{p}^{+}$ and $w_{p}^{-
}=0$), 
and one ``right" branch of level $p$
(using both $Z_{p}^{+}$ and $Z_{p}^{-}$),
to form the ``left" branch structures at level $p+1$
(if $p=q$ we do not continue with forming the $p+1$ level structure,
rather just retain $Z_{p}^{+}$ as the final partition function).
The recursion relations (\ref{seven}) and (\ref{eight}) 
for the restricted partition functions an their derivatives,
respectively, at this step 
take explicit form
\begin{equation}
w_{p+1}^{\pm}=y^{\pm 1}Z_{p}^{+}\left[ Z_{p}^{+}x^{\pm 2} 
+ Z_{p}^{-}\right],\label{nine}
\end{equation}
and
\begin{eqnarray}
&&{\frac{\partial w_{p+1}^{\pm}}{\partial h}}=
y^{\pm 1} \left [ \vbox to 14pt{}
\pm Z^{+}_{p}\left (Z^{+}_{p}{x^{\pm 2}}+Z^{-}_{p}\right )
\right.\nonumber\\
&&\left.+{\frac{\partial Z_{p}^{+}}{\partial h}}
\left (Z^{+}_{p}{x^{\pm 2}}+Z^{-}_{p}\right )+
Z^{+}_{p}
\left (
{\frac{\partial Z_{p}^{+}}{\partial h}}{x^{\pm 2}}+
{\frac{\partial Z_{p}^{-}}{\partial h}}
\right ) \right ].\nonumber\\
\label{ten}
\end{eqnarray}

In the following steps ($p+1<n\leq q$), the ``left" branches 
are combined with (regular) ``right" branches
to form new ``left" branches through general recursion relations 
(\ref{seven}) and (\ref{eight}), while the ``right" branches are all the
time calculated using (\ref{two}) and (\ref{six}).

Evidently, the set of equations (\ref{two}), (\ref{six}-\ref{ten}) is too
complex to be iterated analytically into a closed form expression, for 
arbitrary $p$ and $q$. On the other hand, using algebraic manipulation
software such as Maple or Mathematica, it is a simple exercise to perform these
iterations numerically for arbitrary size and temperature, with arbitrary precision
(depending only on the available computer speed and memory, where present day
personal computers are more then well suited for the job). We have performed
such calculations for various values of $p$, for $q\leq 1024$ (corresponding
to systems of  up to $2^{2^{10}+1}-1\sim 10^{309}$ particles), 
in all cases it was found that fixing {\it any spin} indeed does break the symmetry, 
inducing ordering of enormous
systems in a wide temperature range, while fixing the initial spin does
produce a somewhat stronger effect then fixing a single surface or bulk spin. In fact,
it is found that temperature $T_m$, where magnetization falls off to half of its 
maximum value, monotonously increases with $p$, for any given generation level $q$.
This effect may be attributed to the fact that looking from a given position, 
the average distance 
to all the other spins monotonously decreases as the position shifts from surface
towards the center.
Larger distance from the point of fixing a spin implies a weaker effect,
in accordance with the numerically observed behavior.
\begin{figure}[htp]
\includegraphics[width=3.7in]{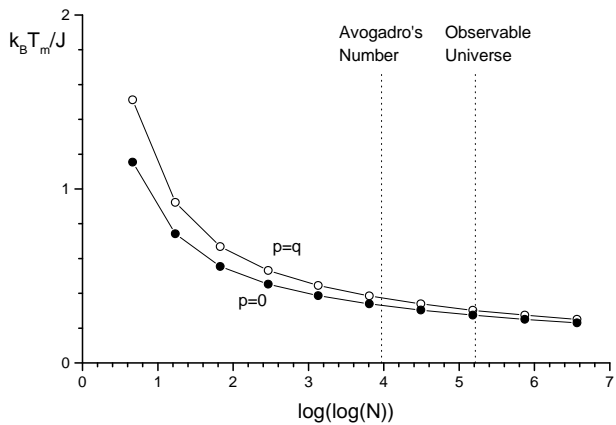}
\caption{
Temperature $T_m$ where magnetization falls to half of its maximum value,
for the Ising system on the Cayley tree, as a function of system size
$N=2^q-1$, on the double logarithmic scale. 
The upper curve (open circles) corresponds to fixing the
initial spin ($p=q$), and the bottom curve (full circles) to fixing
a single surface spin ($p=0$). For all the other values of $0<p<q$,
$T_m$ is found to lie between the two shown curves, monotonously
increasing with $p$.
}
\label{fig3}
\end{figure}
In Fig.~3 we show the comparison of the $T_m$ curves as a function of generation
level $q$ for $p=0$ (full circles), corresponding to fixing a single surface spin,
and $p=q$ (open circles), corresponding to fixing the initial spin.
From the figure, it follows 
that the difference between the values of $T_m$ obtained by fixing
the initial spin and by fixing a single surface spin diminishes with system size,
and that the conclusions about ordering of extremely large systems in a wide
temperature range continues to hold (as well as the comparison of the current model
with the Sierpinski gasket and the chain), if one chooses to break the symmetry
by fixing an arbitrary spin.

In conclusion, as the Cayley tree in itself
(with its infinite dimension and nonzero fraction of surface nodes) represents a 
highly unphysical system in the thermodynamic limit, modeling real world physical 
phenomena (such as, for instance, blood vessels or the nervous system) requires more attention
focus on the finite size behavior. In this work it is shown that the thermodynamic 
limit represents
a questionable approximation for the Ising model on the Cayley tree, as the
magnetic ordering decays extremely slowly with the number of particles. 
As the susceptibility diverges in a wide temperature range in the thermodynamic limit, 
and here we study finite (albeit extremely large) systems,
symmetry breaking is achieved by fixing a single (any one) spin, rather then by
the usual procedure of retaining an infinitesimal field while taking the thermodynamic
limit, and only then taking the limit of zero field.
On a double logarithmic scale with respect to the number of particles, 
we have found that Ising model on the Cayley tree behaves in a fashion
rather similar the behavior of the Ising model on the Sierpinski gasket,
in contrast with the Ising chain. 
However, as opposed to extremely long range
correlations that are present in case of the gasket, 
this behavior may be attributed
to extremely ``wide range" correlations.


\begin{references}

{\bibitem{heimburg}
J. von Heimburg and H. Thomas, J. Phys. C 7 (1974) 3433.}

{\bibitem{matsuda}
H. Matsuda, Prog. Theor. Phys. 51 (1974) 1053.}

{\bibitem{muller}
E. M\" uller-Hartmann and J. Zittartz, Phys. Rev. Lett. 33 (1974) 893.}

\bibitem{morita}
T. Morita and T. Horiguchi, J. Stat. Phys. 26 (1981) 665.

\bibitem{melin}
R. M\' elin, J.C. Angl\' es d'Auriac, P. Chandra \\
and B. Dou\c cot, J. Phys. A 29 (1996) 5773.

\bibitem{stosic}
T. Sto\v si\'c, B.D. Sto\v si\'c and I.P. Fittipaldi, 
J. Mag. Mag. Mater. 177-181 (1998) 185;
ibid. Physica A, 320 (2003) 443.

\bibitem{liu}
S.H. Liu, Phys. Rev. B 32 (1985) 5084.

\bibitem{egarter}
T.P. Eggarter, Phys. Rev. B 9 (1974) 2989.

\end{references}
\end{document}